\documentclass[aps,prd,showpacs,superscriptaddress,groupedaddress]{revtex4}

\usepackage{graphicx,color,bm}

\usepackage{amsmath,amssymb}
\usepackage[colorlinks,linkcolor=blue,citecolor=blue]{hyperref}
\usepackage{epstopdf}
\usepackage{ulem}
\usepackage{array}
\usepackage{verbatim}
\usepackage[utf8]{inputenc}
\usepackage{rotating}

\begin{document}

\title{Observational Constraints on Axion(s) Dark Energy with a Cosmological Constant}

\author{Ruchika}
\email{ruchika@ctp-jamia.res.in}
\affiliation{Centre for Theoretical Physics, Jamia Millia Islamia, New Delhi-110025, India.}

\author{Shahnawaz A. Adil}
\email{shahnawaz188483@st.jmi.ac.in}
\affiliation{ Department of Physics, Jamia Millia Islamia, Delhi-110025, India}

\author{Koushik Dutta}
\email{koushik.physics@gmail.com}
\affiliation{Indian Institute of Science Education And Research Kolkata,
Mohanpur, WB 741 246, India}
\affiliation{Theory Divison, Saha Institute of Nuclear Physics,
HBNI,1/AF Bidhannagar, Kolkata- 700064, India}

\author{Ankan Mukherjee}
\email{ankan.ju@gmail.com}
\affiliation{Department of Physics, Bangabasi College, Kolkata 700009, India.}
\affiliation{Centre for Theoretical Physics, Jamia Millia Islamia, New Delhi-110025, India.}

\author{Anjan A. Sen}
\email{aasen@jmi.ac.in}
\affiliation{Centre for Theoretical Physics, Jamia Millia Islamia, New Delhi-110025, India.}

\vspace{0.5cm}
\vspace{0.5cm}
\pagestyle{myheadings}
\newcommand{\be}{\begin{equation}}
\newcommand{\ee}{\end{equation}}
\newcommand{\bea}{\begin{eqnarray}}
\newcommand{\eea}{\end{eqnarray}}

\begin{abstract}
The present work deals with a dark energy model that has an oscillating scalar field potential along with a cosmological constant (CC). The oscillating part of the potential represents the contribution of a light axion field in the dark energy that has its origin in the String-Axiverse scenario. The model has been confronted with the latest cosmological observations. The results show that a sub-Planckian value of the axion field decay constant is consistent with observational data. Furthermore, in terms of the observational data considered in this work,  the axion model is preferred over the $\Lambda$CDM model in terms of AIC, BIC information criteria as well as  in terms of Bayesian evidence.
The oscillating feature in the scalar field evolution and in the equation of state for the dark energy can be observed for the allowed parameters space. It is also observed that cluster number counts in this axion model are suppressed compared to the $\Lambda$CDM and this suppression is enhanced for the sub-Planckian values for the axion decay constant.   
\end{abstract}
\maketitle

\section{Introduction}

From several cosmological data sets, it is evident that the current Universe is accelerating, and the challenge is to find a mechanism that allows the accelerated expansion of the Universe \cite{Copeland:2006wr}.  Moreover, the solution must be consistent with our acquired knowledge of high energy physics and the laws of gravity. Any explanation in terms of some matter energy component in the Einstein equations is broadly categorised as dark energy. The simplest solution is the existence of a cosmological constant (e.g vacuum energy of a scalar field potential) which acts as a constant energy density source in the expansion equation with an equation of state parameter $w = -1$. Another dynamical solution is the existence of a scalar field(s) $\phi$ which rolls along a flat potential $V(\phi)$ \cite{Ratra:1987rm}, and in this case $w(t)$ being close to $-1$ at the present epoch, to be consistent with the observational data. In both the cases, the value of the cosmological constant or $V(\phi)$ at the present epoch must be tuned to be of the order of $(10^{-3}~eV)^4$. Additionally, for the case of a dynamical scalar field $\phi$, its mass must be of the scale of the current value of the Hubble constant, i.e $m \sim H_0 \leq 10^{-33}$ eV. In this case, the potential must be stable w.r.t quantum corrections which itself is a non-trivial requirement from the point of high energy physics.

 Although, virtually any form of the scalar potential with the above mentioned properties can work as dark energy, an axionic potential with periodic form has several virtues \cite{koushik}. First of all, axion being a pseudo-Nambu Goldstone Boson (pNGB), its potential is generated by breaking of a shift symmetry and therefore its flatness is protected from quantum corrections \cite{frees1990,frieman1995}. Moreover, the axion  being a pseudo-scalar, it easily evades the stringent fifth-constraints due the exchange of $\phi$ quanta \cite{Carroll:1998zi}. When an axion is used as the dark energy field, the field dynamics in late time works as thawing models where equation of state parameter $w$ remains close to $-1$ in early time, and starts to move away from it only in recent time \cite{Caldwell:2005tm}. An axionic potential is parameterised by two scales; $\lambda$ being the overall height of the potential, and it is fixed by the current dark energy density, where as the axionic decay constant $f_1$ determines the periodicity of the potential. The larger the value of $f_1$, more flatter is the potential. For a canonically normalised axion field, the cosmological data  constrain $f_1$ to be close to the Planck scale unless the initial field value is tuned at the top of the potential \cite{Dutta:2006cf, Smer-Barreto:2015pla}. This is typically difficult to arrange in low energy effective field theory of String compactifications, or to motivate from the arguments for gravity being the weakest force \cite{Banks:2003sx,ArkaniHamed:2006dz}. Another possibility with $f_1$ being sub-Planckian is the field being oscillating at the bottom of the potential. But, this case is excluded by observations with high statistical significance \cite{Dutta:2006cf, Adak:2014moa}. One of the important assumption in this discussion is that the value of the cosmological constant is zero, and only one dynamical field with a canonical kinetic energy term contributes to the dark energy density.

 In String Theory, the existence of axions is ubiquitous, and usually the axions appear as complex partners of moduli fields. In the String Axiverse scenario, a large number of axions arise due to compactifications of internal spaces \cite{Arvanitaki:2009fg}. In certain scenario of moduli stablisation, the axion masses of correct order of magnitude required for quintessence field appears naturally. Moreover, internal fluxes contribute to the existence of a cosmological constant whose value can be tuned to be of the same order as required for dark energy \cite{Cicoli:2018kdo}. In this case, acceleration of the Universe can  be driven by complex dynamics of interplay of several axions and a cosmological constant. In the context of String Axiverse scenario, the dynamics of multiple axions contributing to the dynamics of dark energy has been explored in \cite{Kamionkowski:2014zda}. But the analysis is done assuming zero cosmological constant. Cosmological analysis for Axiverse inspired single axion field as dark energy has been done in \cite{Emami:2016mrt}, whereas cosmological implications of ultra-light axion-like fields (not necessarily as dark energy) has been explored in \cite{Poulin:2018dzj}.

 In this work, we would like to analyse the set-up where dark energy dynamics might be governed by multiple axion fields with a cosmological constant \cite{Cicoli:2018kdo}. In particular, we will confront the set-up with the latest available data and will make predictions for structure formation observations. To find an interesting interplay of the dynamical field and the cosmological constant, both the scales for axions and the cosmological constant must be of the same order. Analysis of the above mentioned set-up holds special significance in the 
 context of the latest available data. In particular, seemingly noncompataible value of $H_0$ measurements from the low red-shift observations and the derived value of $H_0$ from the Cosmic Microwave Observations (CMB) observations (See \cite{verde} and references therein) may hint towards nontrivial dynamics of dark energy \cite{collec}. For example, in the presence of a cosmological constant and an axion field, the field may oscillate around its minimum, or may move at the top of the maximum for sub-Planckian axion decay constant. In this work, we do Markov Chain Monte Carlo (MCMC) analysis to find the best fit parameters for the model. We also discuss associated dynamics of the field. 
 
 In Section II, we describe the axion model that we consider in this work; in section III, we describe the background evolution of the model and write down the necessary equations; in Section IV we describe the data used for the analysis and the constraints on the model parameters; in section V, we compare our model with the concordance $\Lambda$CDM model using different information criteria as well as Bayesian evidence; in section VI, we study the spherical collapse formalism in our model to investigate the behaviour of structure formation at the nonlinear scales; finally in section VII, we conclude with the summary of our main results.

\section{Dark energy with Axion potential and a cosmological constant}

An axion potential for a single scalar field $\phi$ is given by,

\be
V(\phi)=\lambda^4\left (1 + \cos{(\phi/f_1)} \right), \label{axion_pot}
\ee
where $f_1$ is the axion decay constant that fixes the periodicity of the potential, and larger the value of $f_1$, more flatter is the potential. For dark energy, the value of $\lambda$ needs to be of the order $10^{-3}$ eV. At the minimum of the potential, potential energy is zero. The single field axionic potential was introduced in the context of inflation by Frees {\it et al} \cite{frees1990} and in the context of dark energy by Frieman {\it et al} \cite{frieman1995}. In the context
of dark energy, the model was confronted with SN data in \cite{Dutta:2006cf}, and it was found that unless the initial value of the field is finely tuned at the top of the potential, the axion decay constant $f_1$ needs to be close to $M_{P}$ . It was also found out that the oscillation of the scalar field at the bottom of the potential is ruled out with high significance (see also \cite{Adak:2014moa}). All the above analysis is done with the following assumptions: (i) the value of the cosmological constant is zero, (ii) there is only one axion field that plays a role in governing the dynamics, (iii) the kinetic energy of the axion field is canonically normalised. 

\par Motivated by the String Axiverse scenario in String Theory \cite{Arvanitaki:2009fg}, there can be several axions 
fields with masses of the order of the current Hubble scale. For the dynamics of these light fields at late time, we can integrate out the relatively heavy d.o.fs and work out with the effective potential. In this case, multi-field axionic potential with canonical kinetic energy term is given by \cite{Cicoli:2018kdo},
\be
V(\phi)=\lambda_0^4+\Sigma_{i=1}^N\lambda_i^4\cos{(\phi_i/f_i)}.
\label{cc_axions_pot}
\ee
where $N$ is the total number of axion fields that can be of order $\mathcal{O}$(100). Each field in the above potential has mass $m_{\phi_i} \sim \lambda_i^2/f_i$. The different decay constants $f_ i$ are for fields $\phi_i$, and $\lambda_0$ is the cosmological constant. The value of $\lambda_0$ in the String Theory is tuned by the fluxes for several realisation of compactification of internal manifold, whereas $\lambda_i$ originates from the non-perturbative effects and sets the mass for the $i$-th axion. In the limit of single axion and $\lambda_i = \lambda_0$, we get back the potential of Eq.~(\ref{axion_pot}). Thus the potential given in Eq.~ (\ref{cc_axions_pot}) introduces the cosmological constant  along with the axionic potential for multiple fields in the dark energy scenario. The value of the potential at the minimum is $V_{min} = \lambda_0^4 - \Sigma_{i = 1}^{N} \lambda_i^4$, whereas the maximum value of the potential is given by $V_{max} = \lambda_0^4 + \Sigma_{i = 1}^{N} \lambda_i^4$. Note that positivity of the value of the potential at the minimum is not a requirement from the observational point of view.

Among several axion fields, the heavier fields will settle down to its minimum earlier during cosmological evolution, and the lightest one will play the crucial role for late time physics \cite{Kamionkowski:2014zda}. At the minima of the heavy fields, the effective potential of the lightest field becomes 
\be
V(\phi)=\lambda^4+\lambda_l^4\cos{(\phi_l/f_l)},
\label{cc_axions_pot1}
\ee
where the contributions from all the heavy fields at their minimum redefines the bare cosmological constant $\lambda^4 = \lambda_0^4 - \Sigma_{i \neq l}^N \lambda_i^4$. Note that $\lambda^4$ can be positive, negative, and even tuned to be zero.  In this paper, we assume $\lambda^4 > 0$. But we should mention that the Universe can still go to de-Sitter or anti de-Sitter phase in future due to the relative magnitudes between $\lambda^4$ and $\lambda_{l}^4$.

\noindent
For an axion field to behave as the dark energy, the slow roll condition must be approximately satisfied:
\begin{equation}
\epsilon_{i} =  \frac{M_{P}^2}{2}\left(\frac{dV/d\phi_i}{V}\right)^2 <1~,
\end{equation}
and for the lightest axion field $\phi_l$ the condition becomes 
\begin{equation}
\epsilon = \frac{1}{2}\left[\left(\frac{\lambda_l}{\lambda}\right)^4  \frac{M_{P}}{f_l}\right]^2 \frac{\sin^2{(\phi_l/f_l)}}{(1 - \left(\lambda_l/\lambda)^4 \cos{(\phi_l/f_l)}\right)^2} < 1.
\end{equation}
It is clear from the above expression that for the usual single field axionic  potential without a cosmological constant when $\lambda_l = \lambda$, the above condition is satisfied if $f_l >  M_{P}$.

If $\lambda_l^4 > \lambda^4$, the potential will have the minimum with a negative potential energy value. In this case the field can roll at the top of the potential where the potential is positive \cite{Kamionkowski:2014zda}. This is akin to the hilltop quintessence \cite{Dutta:2008qn} with the crucial difference that the Universe is going to be at anti-de Sitter state at far future. In the case of multiple ultralight axion fields in hand, it is not unnatural that some fields will have initial field values at the top of the potential. But it is important that the lightest field is among those fields for the set-up to work. 

On the other hand, if $\lambda \gtrsim \lambda_l$, the set-up for quasi-natural quintessence can be realised for sub-Planckian value of $f_l$ \cite{Cicoli:2018kdo}. In this case, the most of the dark energy contributions come from the cosmological constant with axionic part contributes as modulations. The field rolls neither at the peak of the potential nor at the bottom. Another interesting possibility emerges when the value of $\lambda$ is nearly tuned to the value of current dark energy density, and the axionic field oscillates around the minimum of the potential. In this case, oscillations of the scalar field is reflected in the oscillating equation of state parameter $w$. Note that because of the existence of $\lambda$, the oscillating scalar field does not behave as pressure less matter. 

In this work, we analyse these cases in light of the latest available cosmological data.

\section{System of equations for a dynamical dark energy}
\label{dynamicalsys}

The cosmological principle says that the universe is homogeneous and isotropic at cosmological scale. It is the basic assumption to study the cosmos at background level. The evolution of background cosmology is governed by Friedmann equations. In a spatially flat FRW universe, Friedmann equations are given in terms of Hubble parameter ($H$) and its time derivative as,
\begin{equation}
3H^2 = \kappa^2\rho_{tot},
\label{friedmann1}
\end{equation}
\begin{equation}
2\dot{H}+3H^2 = -\kappa^2p_{tot},
\label{friedmann2}
\end{equation}
where the overhead `dot' denotes the derivative with respect to time,  $\kappa^2 = 8 \pi G = M_{P}^{-2}$. The  $\rho_{tot}$ and $p_{tot}$ are respectively the total energy density and pressure of all the components present in the energy budget of the universe. The Hubble parameter is defined as, $H=\dot{a}/a$ where $a$, the {\it scale factor}, takes care of the time evolution of the spatial separation between two points at cosmological length scale. The contribution to the energy density comes from matter (dark matter and baryonic matter), radiation and the dark energy, thus $\rho_{tot}=\rho_m$+$\rho_r+\rho_{de}$. In the present model, dark energy consists of a cosmological constant and contribution from one light axion field.  The axionic potential with cosmological constant, given in equation (\ref{cc_axions_pot1}), can be rewritten as,
\begin{equation}
V(\phi) = V_0 [1 + p \cos{(\phi/f_1)}],
\label{ax_pottetial_2}
\end{equation} 
where $V_0=\lambda^4$, $p=(\lambda_l/\lambda)^4$, and $\phi$ is the lightest field. Note that, $p = 0$ corresponds to the existence of only the cosmological constant, whereas for $p = 1$, the potential reduces to the case of single field axionic potential without any cosmological constant. The energy density and pressure like contribution of a scalar field dark energy are $\rho_{\phi}=\frac{1}{2}\dot{\phi}^2+V(\phi)$ and $p_{\phi}=\frac{1}{2}\dot{\phi}^2-V(\phi)$ respectively. The equation of state parameter for the scalar field dark energyis expressed as,
\be
w=\frac{\dot{\phi}^2-2V(\phi)}{\dot{\phi}^2+2V(\phi)}.
\ee
  The conservation equation of the dark energy density ($\rho_{\phi}$) is,
\be
\dot{\rho}_{\phi}+3H(1+w)\rho_{\phi}=0,
\label{de_conserve}
\ee
which yields the Klein-Gordon equation for the scalar field in the expanding background,
\be
\ddot{\phi}+3H\dot{\phi}+\frac{dV}{d\phi}=0.
\label{scalKlineGordon}
\ee
The conservation equations for pressureless dust matter and radiation are respectively given as,
\be
\dot{\rho}_m+3H\rho_m=0,
\label{mat_conserve}
\ee
\be
\dot{\rho}_r+4H\rho_r=0.
\label{rad_conserve}
\ee
Dimensionless density parameters are defined as $\Omega_{i}=\frac{\rho_i}{3H^2/\kappa^2}$. In a spatially flat universe, they are connected as, 
\be
\Omega_{m}+\Omega_{r}+\Omega_{\phi}=1.
\ee
Next, we define few dynamical variables $x$, $y$ and $\Lambda$ \cite{senscherrer} as,
\begin{eqnarray}
x = \kappa\phi' / \sqrt{6} \quad ; \quad \quad y = \sqrt{\kappa^2V(\phi)/3H^2} \quad ; \quad \Lambda = - \frac{1}{V} \frac{dV}{d \phi}
\end{eqnarray}
where the `prime' denotes derivative with respect to $\ln{a}$. 
The scalar field dark energy density parameter can be expressed in terms of $x$ and $y$ as, $\Omega_{\phi} = x^2 + y^2$. We define another dimensionless quantity $\gamma$ as,
\begin{equation}
\gamma = 1 + w = \frac{2 x^2}{x^2 + y^2}.
\end{equation}
The coupled dynamical system equations of the dimensionless variables $\gamma$, $\Omega_{\phi}$  and $\Lambda$ can now be written as, 
\begin{eqnarray}
\gamma' = -3 \gamma(2-\gamma)+\Lambda(2-\gamma)\sqrt{3\gamma\Omega_{\phi}},\\
\label{gamma_eq}
\Omega' _{\phi}= 3(1-\gamma)\Omega_{\phi}(1-\Omega_{\phi})+\Omega_{\phi}\Omega_{r},\\
\label{Omphi_eq}
\Lambda' = -\sqrt{3}\Lambda^2(\Gamma-1)\sqrt{\gamma\Omega_{\phi}},
\label{lambda_eq}
\end{eqnarray}
where 
\begin{equation}
\Gamma = \frac{V \frac{d^2V}{d\phi^2}} {(\frac{dV}{d \phi})^2 }
=  \frac{-p^2 + \sqrt{(p^2-1)(\Lambda f_1)^2 + p^2} }{(p^2-1)(\Lambda f_1)^2}. 
\end{equation}
The solution of these autonomous system of equations  are obtained numerically. 
It is always preferable to work in terms of observable quantities as it is easier to assign priors for the statistical analysis. To solve the above set of autonomous system of equations, the values of the parameters are initialized at redshift z=1100. Assuming thawing class of evolution for the scalar field \cite{Caldwell:2005tm}, the initial value of $\gamma$ is taken to be $\gamma=1+w= 0.0001$ or $w$ near to $-1$. Initial values of the other quantities, namely $\Omega_{\phi}$ and $\Lambda$ are kept as free parameters,  $\Omega_{\phi i}$ and $\Lambda_{i}$. 

To incorporate the linear growth rate data ($f\sigma_8$) from large scale structure survey, the evolution of matter density contrast at linear regime is needed to be studied.  The matter density contrast is defined as $\delta=\delta\rho_m/\rho_m$, where $\delta\rho_m$ is the deviation from background matter density. The linear equation of the matter density contrast is given as, 
\begin{equation}
\ddot{\delta}+2H\dot{\delta}=4\pi G\rho_m\delta.
\label{delta_linear}
\end{equation}
Equation (\ref{delta_linear}) can be written in terms of the dimensionless variables as,
\begin{equation}
\delta'' = -(0.5 - 0.5\Omega_r-1.5\Omega_{\phi}(\gamma-1))\delta'+1.5(1-\Omega_{\phi}-\Omega_{r})\delta.
\label{delta_linDimensionless}
\end{equation}
The cosmological evolution governed by the autonomous set of equations (eqns (17) to (18)) and the equation of linear matter density contrast (eqn. \ref{delta_linDimensionless}) are confronted with cosmological observations to obtain the constraints on the model as well as cosmological parameters. The observational data sets, utilized in the present statistical analysis, and the observational constraints are discussed in the following section. 


\section{Statistical analysis of the model and constraints on cosmological parameters}
\label{Obconstraint}

The dark energy model with axionic potential and cosmological constant, discussed in the present context, has been confronted with latest cosmological observations. The observational data sets, utilized to constraint the model are briefly discussed in the following.

\begin{itemize}

\item {Observational measurements of Hubble parameter as a function of redshift using cosmic chronometers (CC) as compiled by Gomez-Valent and Amendola \cite{Gomez-Valent:2018hwc}, has been utilized in the present analysis.}

\item {The latest SH0ES measurement of $H_0$, denoted as R21, is also included in the analysis \cite{Riess:2021arx}.}

\item {The distance modulus measurement of type Ia supernovae compiled in the latest Pantheon sample \cite{Gomez-Valent:2018hwc} (here- after “SNe” data) in terms of $H(z)/H_0$ has been incorporated in the analysis.}

\item {The CMB shift parameter and acoustic scale measurements from Planck-2018 results   \cite{Chen:2018dbv}.  The method to reduce the information from a full CMB likelihood function to fewer number of parameters were first discussed by Koskowsky et al. \cite{koskow} and Wang \& Mukherjee \cite{wangmukh}. In the Planck 2015 analysis for cosmological results, the compressed likelihood involving the shift parameter $R$ and acoustic scale $l_{a}$ were obtained (see  Ade et al. \cite{planck2015} for the detail method for obtaining the compressed likelihood). It was argued that this compressed likelihood can be used jointly with other likelihood functions to constrain dark energy models which are smooth which is what we are considering in our work. In this work, we have used the compressed likelihood for Planck-2018 obtained by Chen et al \cite{Chen:2018dbv}.}

\item {Isotropic baryon acoustic oscillation (BAO) measurements by 6dF survey (z =0.106) \cite{6df}, SDSS-MGS survey (z = 0.15) \cite{Ross:2014qpa} as well as by BOSS quasar clustering (z=1.52) \cite{Ata:2017dya}. We also consider anisotropic BAO measurements by BOSS DR12 at z = 0.38, 0.51, 0.61 \cite{Alam:2016hwk}. Also, we considered BAO measurement by BOSS-DR12 using Lyman-$\alpha$ samples at z = 2.4 \cite{Bourboux:2017cbm} in the present analysis.}

\item {We also incorporate the angular diameter distances measured using water megamasers under the Megamaser Cosmology Project at redshifts z = 0.0116, 0.0340, 0.0277 for Megamasers UGC 3789, NGC 6264 and NGC 5765b respectively \cite{Reid:2008nm,reid,kauC,Gao:2015tqd}, hereafter written as “MASERS”.}

\item {Strong lensing time-delay measurements by H0LiCOW
experiment (TDSL) by Bonvin et al. (2017) \cite{Bonvin:2016crt} for three lens systems are also utilized in the present context.}

\item {We use the measurement of $f\sigma_{8}$ by various galaxy surveys as compiled by Basikalos et al \cite{basikalos}}

\end{itemize}

We do a Markov Chain Monte Carlo (MCMC) analysis using the observational data to constrain the model parameters and evolution of cosmological quantities.  The analysis is carried out using the {\footnotesize PYTHON} implementation of MCMC sampler, the {\footnotesize EMCEE} hammer introduced by Foreman-Mackey {\it et al} \cite{emcee} which is widely used in cosmological data analysis. The parameter space, analyzed in the present context consists of ($h,r_d,\sigma_8,\Omega_{\phi i},\Lambda_i,f_1,p$). We consider both the $p  > 1$ (in which case, the Universe is asymptotically anti de-Sitter in future) as well as $ p < 1$ (in which case, the Universe is asymptotically de-Sitter in future) cases.

We have used uniform priors for all the cosmological parameters used in our model. The Hubble parameter at present ($z=0$) in our subsequent calculations is assumed to be $H_{0} = 100 h km/s/Mpc$, thus define the dimensionless parameter $h$. We fix $\Omega_{r0}$ at present to be  $5\times 10^{-5}$ in our subsequent calculations. Also the parameter $\Omega_{b0}h^2$ appears while incorporating the CMB compressed likelihood, discussed above, hence we also vary this parameter while fitting the data. The priors corresponding to each parameter is listed in Table I.

\begin{table*}[h!]
\centering
\begin{tabular}{|c|l|}
\hline
Quantity&Prior\\
\hline\hline
$\Omega_{\phi i}/10^{-9}$  & $[1,3]$ \\
\hline
$r_d$&$[130,170]$ Mpc\\
\hline
 $\Lambda_{i}$    &  $[10^{-6}, 1.5]$\\ 
 \hline
 $\Omega_{b0}h^2$  &  $ [0.005, 0.1]$   \\
 \hline
 $f_1$  &  $[0.01, 1.2]$ $\sqrt{3}M_p$   \\
  \hline
 $p$  &  $[0.1, 0.995]$  / $[1.04, 1.15]$\\
  \hline
 $h$  &  $[0.5, 0.9 ]$ \\
  \hline
 $\sigma_8$  &  $[0.6, 1.0]$\\
\hline
\end{tabular}
\caption{The ranges of the uniform prior of the parameters used in the present MCMC analysis.}
\label{priortab}
\end{table*}

\begin{figure}[htb!]
\begin{center}
\resizebox{500pt}{500pt}{\includegraphics{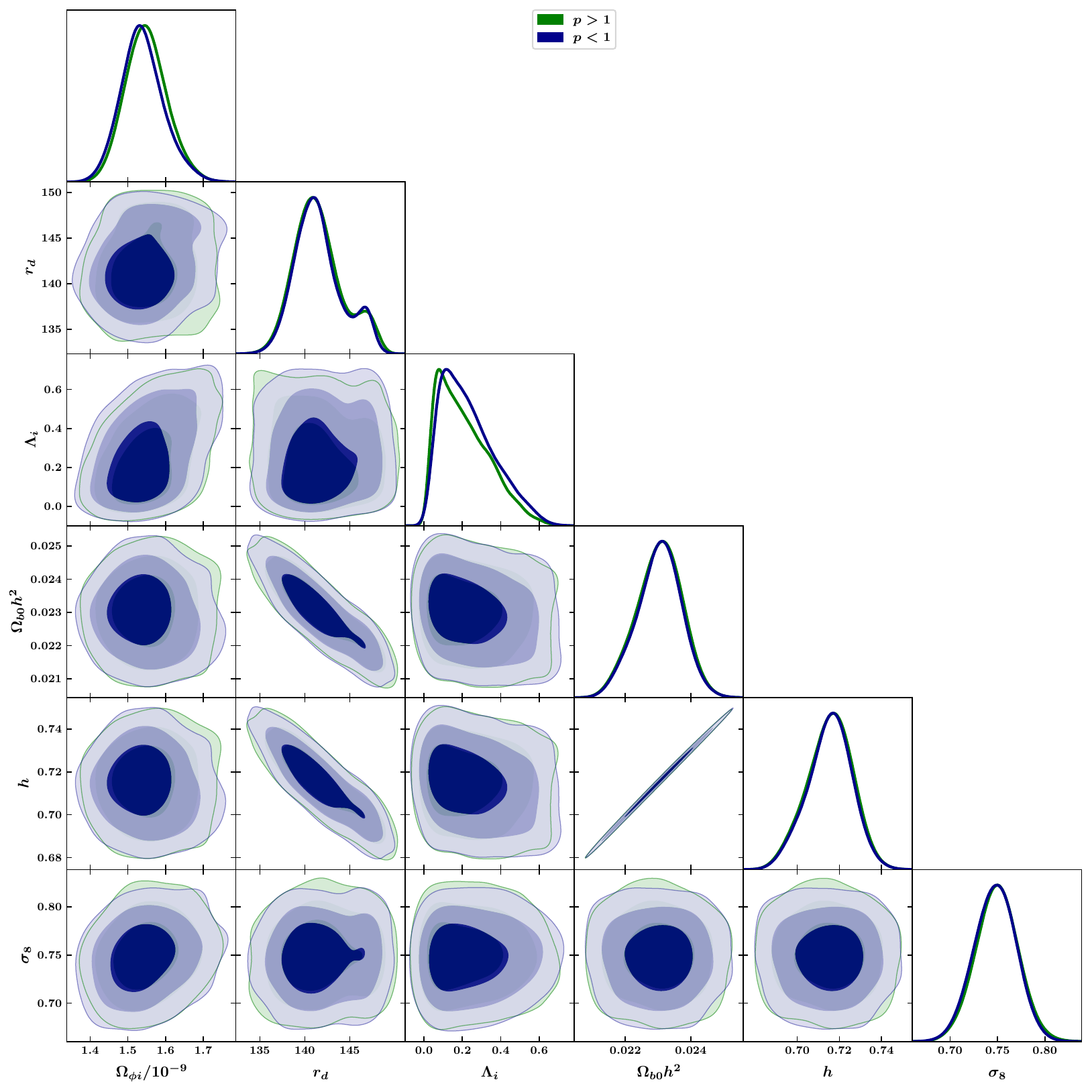}}
\end{center}
\caption{{\small Marginalized posterior distribution of the set of parameters ($\Omega_{\phi i}/10^{-9},r_d,\Lambda_i,\Omega_{b0}h^{2},h,\sigma_8$) and corresponding 2D confidence contours for both $p < 1$ and $p > 1$ cases, obtained from the MCMC analysis for the present model utilizing all the data sets mentioned in section
\ref{Obconstraint}.}}
\label{posterior_2Dcontours}
\end{figure}

\begin{table}[hbt!]
\begin{center}
\scalebox{0.97}{
\renewcommand{\arraystretch}{2.0}
\begin{tabular}{|c |c |c |c|c|c|c|c|c|c|    } 
 \hline
Parameters & $\Omega_{\phi i}/10^{-9}$  &  $r_d$(Mpc) &  $\Lambda_{i}$    & $\Omega_{b0}h^2$  &  $f_1$ $[M_p]$  &   $p$  &   $h$  &  $\sigma_8$ & $\Omega_{m0}$\\ 
 \hline
Constraints (p$<$1) & $1.539^{+0.05}_{-0.061}$  & $141.6^{+1.8}_{-3.3}$  & $0.23^{+0.083}_{-0.18}$  & $0.023^{+0.0007}_{-0.0006}$ 
 & $0.71^{+0.27}_{-0.27} $  
& $0.57^{+0.30}_{-0.23} $  
 & $0.715^{+0.012}_{-0.009}$  
& $0.747^{+0.022}_{-0.022} $
& $0.298 ^{+0.007}_{-0.007}$  \\
 \hline
 Constraints (p$>$1) & $1.548^{+0.052}_{-0.060}$  & $141.6^{+1.9}_{--3.4}$  & $0.209^{+0.076}_{-0.18}$  & $0.023 ^{+0.00076}_{-0.00065}$ 
 & $0.77^{+0.38}_{-0.21} $  
& $1.094^{+0.032}_{-0.032} $  
 & $0.715^{+0.012}_{-0.009}$  
& $0.749^{+0.022}_{-0.022} $
& $0.298^{+0.007}_{-0.007}$  \\
\hline
\end{tabular}
}
\caption{{\small The parameter values, obtained in the MCMC analysis along with the derived parameter $\Omega_{m0}$ combining all the data sets discussed in section \ref{Obconstraint}, are presented along with the associated 1$\sigma$ uncertainty. }} 
\end{center}
\label{besttab}
\end{table}

The constraint obtained for different parameters are shown in Table II. The posterior probability distributions of different parameters and the corresponding 2D confidence contours are shown in Figure \ref{posterior_2Dcontours}.  As one can see, the constraints on different cosmological parameters are similar for both $p < 1$ and $p > 1$ cases. The central value for $f_{1}$ is sub-Planckian for $ p > 1$ whereas it is super-Planckian for $ p <1$ case, but within $1\sigma$ confidence limit (lower) itself, sub-Planckian values for the axion decay constant  $f_{1}$ is consistent with the observed data for both $ p >1$ and $p < 1$ cases. Given the constraints on the parameters mentioned in Table I and Table II, one can derive constraint on the initial value of the axion field $\phi_{i}$ which is $(0.9 \pm 0.44) M_{p}$ (for $p<1$) and $(0.56 \pm 0.31) M_{p}$ for ($p>1$). Moreover one can also calculate the mass of the axion field $m_{\phi}$ which is $\approx 10^{-35}$ eV.

 We should also mention that in our data analysis, we have used the SH0ES prior \cite{Riess:2021arx} while incorporating the Pantheon data for SnIa. But we have checked that the results does not change if instead we use the prior for the intrinsic magnitudes of the SnIa as discussed in \cite{Efstathiou}.

\begin{figure}[hbt!]
\begin{center}
\includegraphics[angle=0, width=0.4\textwidth]{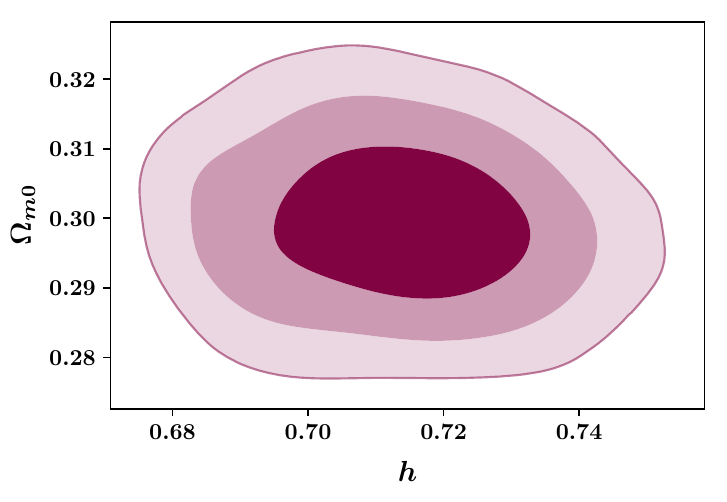}
\includegraphics[angle=0, width=0.4\textwidth]{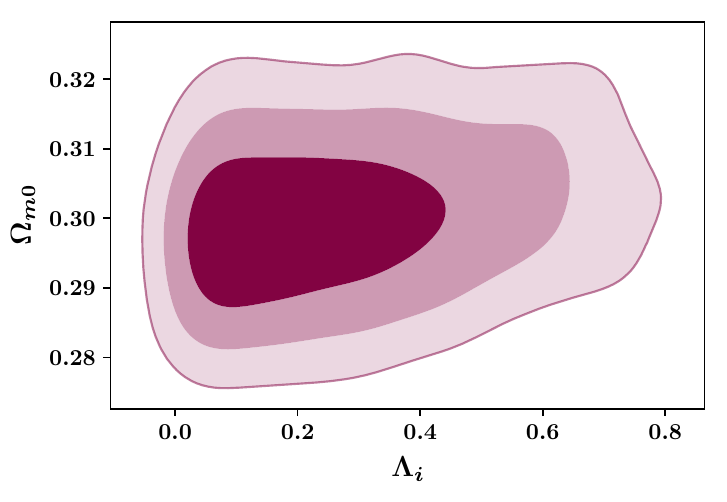}
\end{center}
\caption{\small Confidence contours on different 2D parameter spaces ($h,\Omega_{m0}$) and ( $\Omega_{m0},\Lambda_i$)) (for $p <1$ case) .}
\label{OmContours}
\end{figure}

\begin{figure}[hbt!]
\begin{center}
\includegraphics[angle=0, width=0.55\textwidth]{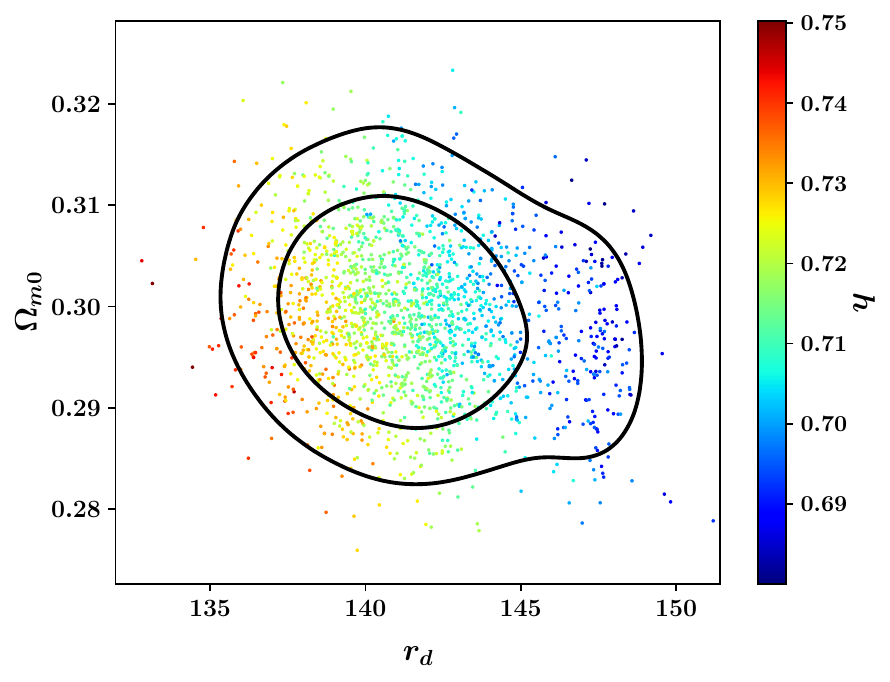}
\end{center}
\caption{{\small Confidence contours on the 2D parameters space consist of $\Omega_{m0}$ and $r_d$ (in Mpc). The $h$ values are indicated by colour index. (for $p <1$ case)}}
\label{Om_rd}
\end{figure}

With the result that the constraints on the cosmological parameters are insensitive to whether $ p >1$ or $ p < 1$, we continue our further study with the $p < 1$ case where the Universe is asymptotically de-Sitter in future. So all the further results below are for $p < 1$ case only and we emphasize there is no change in these results for $ p >1$ case. The central value of $p = 0.577$ corresponds to the case where the total dark energy contributions come from both the cosmological constant and the axion contributions. In fact, in this case, the axion contribution adds small modulation on top of the cosmological constant. However, it is important to remember that both the parameters $f_1$ and $p$ have large error bars. Considering $2$-$\sigma$ error bars, the data can not distinguish between a pure cosmological constant and a pure axionic potential.  
Figure \ref{OmContours} shows the 2D confidence contours of the present matter density parameter $\Omega_{m0}$ with other parameters such as $h$ and $\Lambda_i$. $\Omega_{m0}$ shows almost no correlation with $h$ and $\Lambda_i$. Figure \ref{Om_rd} presents the 2D contour plot of ($\Omega_{m0}, r_d$) along with the corresponding $h$ values indicated by the colour coding. We get slightly lower value for $r_{d}$ due to comparatively higher value for $H_{0}$ which is consistent with the fact that the combination $r_{d}H_{0}$ is fixed by the BAO measurements. The constraint with $2\sigma$ error bar on present day value for the equation of state of this axion field is $w_{0} = -0.982^{+0.055}_{-0.017}$.

\begin{figure}[hbt!]
\begin{center}
\includegraphics[angle=0, width=0.55\textwidth]{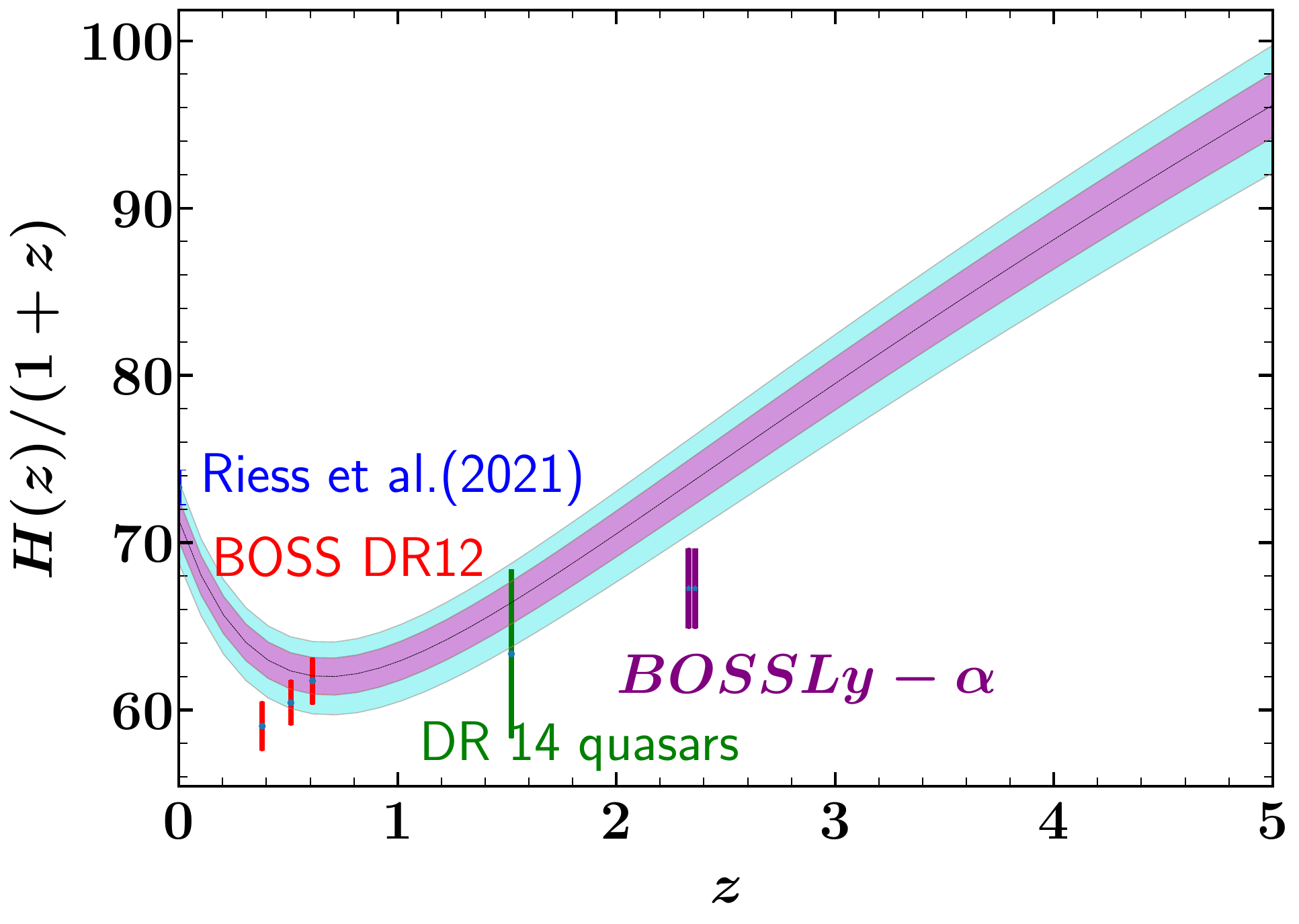}
\end{center}
\caption{{\small Plot of $H(z)/(1+z)$ (km/s/Mpc.) In this plot, the black line is best fit for the present model and inner and outer shaded regions are for 1$\sigma$ and 2$\sigma$ respectively. (for $p <1$ case) }}
\label{hzandw}
\end{figure}

\noindent
Figure \ref{hzandw} shows the behaviour of reconstructed $H(z)/(1+z)$. Our reconstructed value of $H(z)/(1+z)$ at $z=2.36$ is $73.76 \pm 1.44$ km/s/Mpc which is more than $2 \sigma$ away from observational BOSS Ly-$\alpha$ data ($67.26 \pm 2.38$ km/s/Mpc). Except for this individual result, our reconstructed $H(z)/(1+z)$ is consistent with combined BAO measurements for $H(z)/(1+z)$.

\vspace{10mm}


\begin{figure}[hbt!]
\begin{center}
\resizebox{180pt}{120pt}{\includegraphics{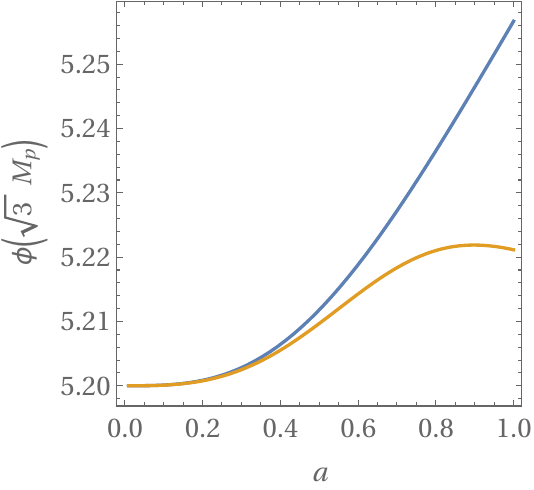}} 
\hspace{4mm}
\resizebox{200pt}{120pt}{\includegraphics{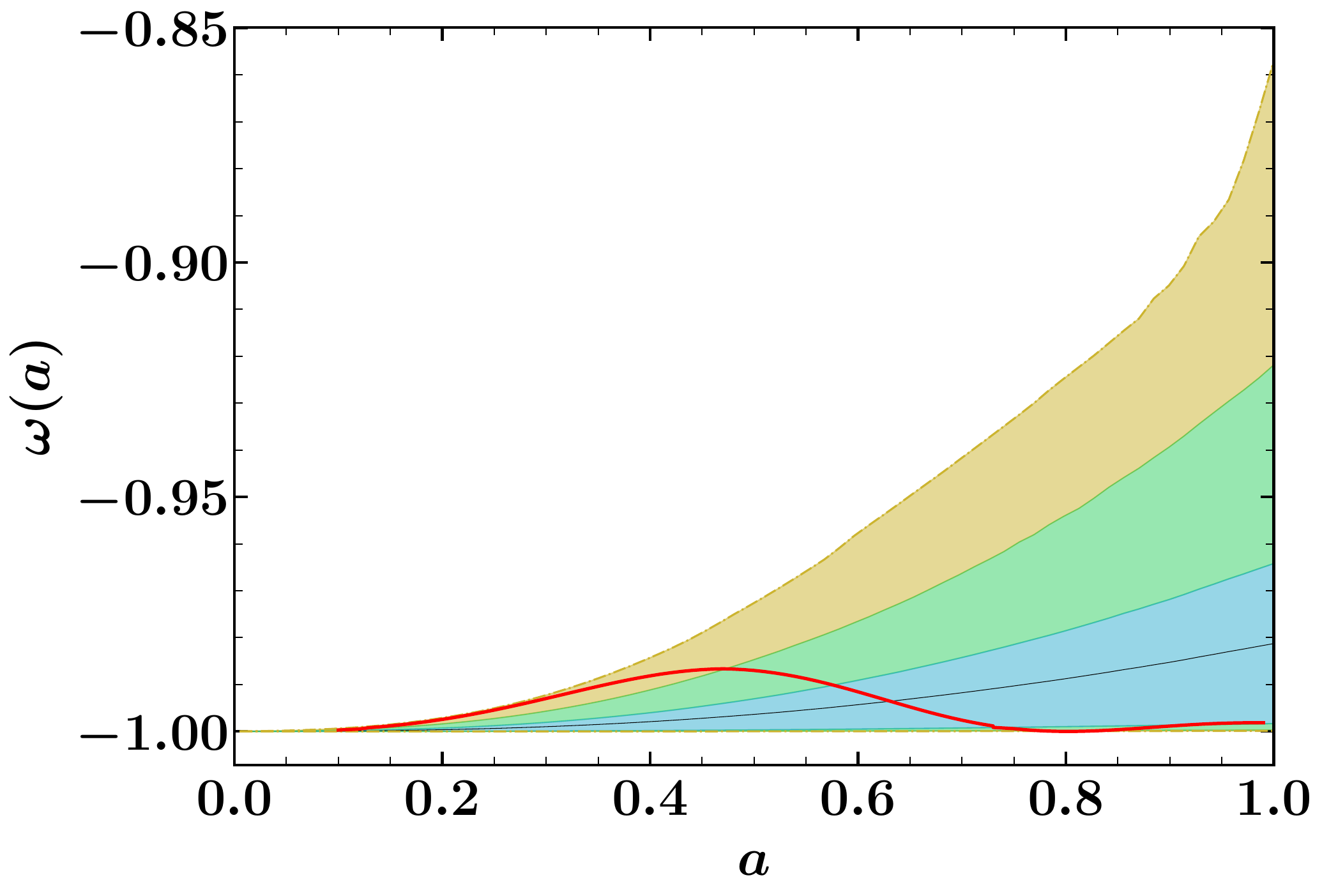}} \\
\vspace{5mm}
\includegraphics[angle=0, width=0.39\textwidth]{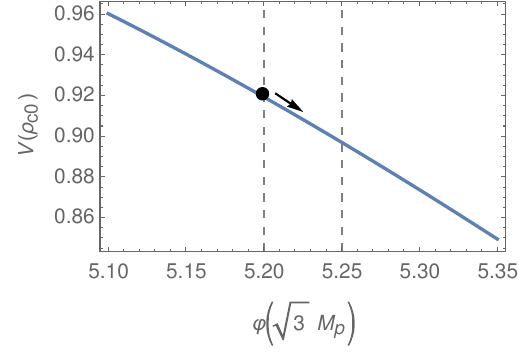}
\includegraphics[angle=0, width=0.40\textwidth]{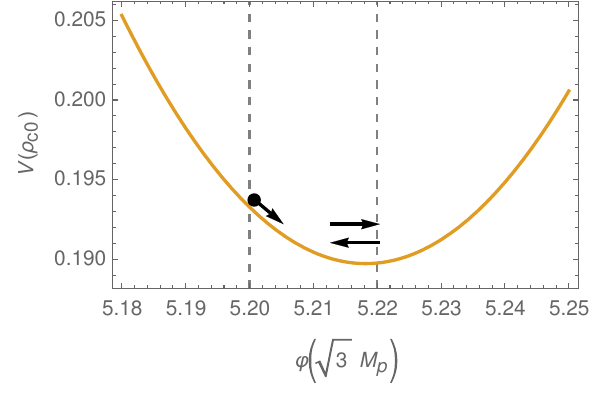}
\end{center}
\caption{Evolution of the scalar field (upper left panel) and equation of state parameter $w(z)$ (upper right panel) and the potential $V(\phi)$ in lower panel. 
In the top-right panel, the reconstructed region for $w(z)$ is shown with best fit (black line) as well as the $1\sigma$ (darker) and $2\sigma$ (lighter) region. The red line is a typical oscillatory behaviour for parameter choice $f_1=0.262 M_p$ and $p=0.721$ within $2\sigma$. The corresponding best fit curves for $\phi$ and $V(\phi)$ are drawn in blue in other panels. The other curves (yellow curves) in  $\phi$ and $V(\phi)$ plots are obtained for $f_1=0.262 M_p$ and $p=0.721$ for which the $w(z)$ shows oscillations at $2\sigma$. The region of rolling as well as oscillation for the scalar field is indicated by the dashed bars. (for $p <1$ case)}
\label{phiw_oscillation}
\end{figure}

\noindent
\noindent
The evolution of the scalar field $\phi$ and the dark energy equation of state parameter $w$ as a function of scale factor are shown in figure \ref{phiw_oscillation}. The best fit curve of $\phi(a)$ and $w(a)$ hardly show any oscillating nature. But for certain values of the parameter $p$ and $f_1$, selected from the 2$\sigma$ confidence regions of the these parameters (as mentioned in the caption), oscillating features in $\phi(a)$ and $w(a)$ are observed. This is definitely due to the interplay of the cosmological constant and the axion potential. Note that for a pure axion potential without a cosmological constant, the oscillation at the bottom of the potential is excluded with high significance \cite{Dutta:2006cf, Adak:2014moa}. Top panels of figure \ref{phiw_oscillation} shows evolution of the scalar field and the dark energy equation of state. The bottom panel of figures \ref{phiw_oscillation} shows the region in the potential where the scalar field evolves as shown in the top panel. In the figure for the equation of state $w(z)$, we show the reconstructed behaviour for $w(z)$ (with $1\sigma$ and $2\sigma$ region) as well as an oscillatory behaviour which occurs for certain choice of the parameters within $2\sigma$ region.The region of oscillation of $\phi$ is also indicated in lower panels of figure \ref{phiw_oscillation}. This figures confirms the fact that within $2\sigma$ confidence interval, the data allows the oscillatory behaviour, both in the scalar field evolution as well as in the dark energy equation of state.

\begin{table*}
 \centering
 \caption{The $\chi^{2}_{min}$ obtained for  the axion model (p $<$ 1 and p $>$ 1) and the concordance $\Lambda$CDM for various data combinations.
}
\begin{tabular}{c  c  c }
\hline \hline
Data & Model & $\chi^2_{\rm min}$\\
\hline
& p $<$ 1 & $86.33$  \\
\hline
All Data & p $>$ 1 & $86.322$    \\
\hline
& $\Lambda CDM$ & $104.874$  \\
\hline
\hline
& p $<$ 1 & $81.297$  \\
\hline
All Data -$H_0$ &p $>$ 1& $81.59$   \\
\hline
& $\Lambda CDM$ & $90.58$  \\
\hline 
\hline
& p $<$ 1 & $84.69$  \\
\hline
All Data -CMB &p $>$ 1& $84.623$   \\
\hline
& $\Lambda CDM$ & $98.49$  \\
\hline 
\hline
& p $<$ 1 & $80.09$  \\
\hline
All Data -CMB -$H_{0}$&p $>$ 1& $80.34$   \\
\hline
& $\Lambda CDM$ & $88.927$  \\
\hline 
\hline
\end{tabular}
\end{table*}

\section{Model Selection}

In this section, we compare the axion model for both $p<1$ ( asymptotic de-Sitter) and $p>1$ ( asymptotic anti de-Sitter) scenarios with the flat $\Lambda$CDM model which is by far the simplest dark energy model that is consistent with majority of cosmological data. To do such comparison, we use three statistical estimators that are widely used for comparing dark energy models.

The first two are Akaike Information Criterion (AIC ) and Bayesian Information Criterion (BIC). For these, we have two parameters ``AIC" and ``BIC" that depend on the $\chi^2_{min}$, number of parameters in the model $d$ and the number of data points $N$. They are defined as:

\begin{eqnarray}
AIC &=& \chi^{2}_{min} + 2 d\nonumber\\
BIC &=& \chi^{2}_{min} + d \ln(N).
\end{eqnarray}

\noindent
With that definitions, while comparing between two models, model with lower AIC or BIC is preferred. So for AIC, to see whether the axion model with more number of parameters are preferred over $\Lambda$CDM, one should have at least an improvement of $\Delta\chi^2 \geq 2 \Delta d$ for axion model to be preferred over  $\Lambda$CDM. For BIC, models with more number of parameters, compared to  $\Lambda$CDM, are penalised further due to the $\ln(N)$ factor.

The third estimator to compare models is the Bayesian Evidence $(\cal Z)$ or the marginalised global likelihood. It is defined as:

\begin{equation}
{\cal Z} = \int d\Theta  {\cal P} (\Theta|{\cal M})  {\cal P} ({\cal D}|\Theta,{\cal M}).
 \end{equation}
 
 \noindent
 Here ${\cal M}$ , ${\cal D}$ and $\Theta$ represent model, data and parameters in the model respectively. The first term in the integrand represents the prior probability distribution and the second term is the likelihood function. According to the Jeffrey's scale  \cite{jeffrey}), for two models, $\Delta ln {\cal Z} > 2.5$ gives a strong evidence for the model with higher ${\cal Z}$ compared to the model with lower ${\cal Z}$.
 
 \begin{table*}
 \centering
 \caption{The AIC/BIC obtained for  the axion model (p $<$ 1 and p $>$ 1) and the concordance $\Lambda$CDM for combination of all data.
}
\begin{tabular}{c  c  c c c  c c c c c}
\hline \hline
 Model & $\chi^2_{\rm min}$ & $d$ & $N$ & ${\rm AIC}$ & ${\rm BIC}$ & $\Delta {\rm AIC}$ & $\Delta {\rm BIC}$ \\
\hline
p $<$ 1 & $86.33$ & $8$ & $70$ &  $102.33$ & $120.318$ & $0.0$ & $0.0$  \\
\hline
 p $>$ 1 & $86.322$ & $8$ & $70$ &  $102.322$ & $120.31$ & $0.008$ &$0.008$    \\
\hline
 $\Lambda CDM$ & $104.874$ & $5$ & $70$  & $114.874$& $ 126.1165$ & $12.544$ &$5.7985$ \\
\hline
\hline
\end{tabular}\label{infop}
\end{table*}

In Table III, we show the $\chi^{2}_{min}$ for different data combinations for two different cases ($p<1$ and $p>1$) as well as for $\Lambda$CDM model. In Table IV we show the
AIC and BIC values obtained for these models for all  data combinations.  As one can see, with three extra parameters in axion model compared to $\Lambda$CDM,  
there is always an improvement in $\chi^{2}_{min}$ for the axion model (for both  $p<1$ and $p>1$) compared to $\Lambda$CDM.  With all the data, this improvement is sufficiently large resulting a large reduction in AIC parameter ($\Delta AIC > 10$ ) in axion model (for both  $p<1$ and $p>1$) compared to $\Lambda$CDM. This shows that axion model is very strongly preferred over $\Lambda$CDM in term of AIC criteria. In the case of BIC also, the axion model is preferred over $\Lambda$CDM (between 2 and 6).

\begin{table}[hbt!]
\begin{center}
\begin{tabular}{|c |c |c |c |   } 
 \hline
Models & $\Lambda$CDM  &  Axion Model$ (p < 1)$ & Axion Model $(p > 1)$ \\ 
 \hline
Ln \textit{Z} & $ -68.877$  & $-65.81$  & $ -66.656$   \\
 \hline

\end{tabular}

\caption{\small  Log Evidence\textit{(Z)} for $\Lambda$CDM Vs Axion Model($p<1,p>1$) }
\end{center}, 
\label{tabevd}
\end{table}

Next we compare the axion model with $\Lambda$CDM using Bayesian Evidence estimator  MultiNest \cite{feroz}. For this purpose we have used python interface to it called PyMultinest \cite{Buchner}. As shown for AIC/BIC estimators, local $H_{0}$ measurement does not play a big role for this selection criteria. Hence we quote the results for the Bayesian Evidence considering all the data mentioned in section IV. The prior for the parameters are mentioned in Table I. They are sufficiently wide prior. And we also verify that our results are stable for wider priors. The results are shown in Table V.  We show the Bayesian Evidences for the three cases: $\Lambda$CDM, $p >1$ and $p < 1$.  It is evident from these numbers that both $p < 1$ and $p > 1$ cases have strong Bayesian evidences ($\Delta ln {\cal Z} > 2.5$ in Jeffrey's scale \cite{jeffrey}) compared to the $\Lambda$CDM model. On the other hand, there is no preference between $p < 1$ and $p > 1$ cases as they have similar Bayesian evidences.

To summarise, both AIC and BIC estimators as well as Bayesian evidence show that axion model is preferred over $\Lambda$CDM. AIC estimator and Bayesian Evidence  in particular, very strongly prefer axion model over $\Lambda$CDM. And this is true for both asymptotic de-Sitter and anti de-Sitter scenarios.

\section{Spherical collapse of matter density perturbation and galaxy cluster number count}

In this section, we have emphasised on the collapse of matter density perturbation and the formation of large scale structure in the universe for present model where dark energy contains axion field along with cosmological constant. We have assumed a spherical collapse scenario of the matter over density. Spherical collapse model \cite{Gunn:1972sv,Liddle:1993fq,Nunes:2004wn} is the simplest approach to trace the evolution of matter over density in the non-linear regime. The idea is that an over-dense region would expand with the Hubble expansion but also gather mass due to gravitational attraction. At some time, it will 
collapse depending on the length scale. The density contrast of cold dark matter is defined as  $\delta=\delta\rho_m/\rho_m<<1$. The non-linear evolution of the matter  density contrast is governed by the equation given as,
\be
\ddot{\delta}+2H\dot{\delta}-\frac{\kappa^2}{2}\rho_m\delta(1+\delta)-\frac{4}{3}\frac{\dot{\delta}^2}{(1+\delta)}=0.
\label{deleq}
\ee
At linear regime, equation (\ref{deleq}) is simplified as,
\be
\ddot{\delta}+2H\dot{\delta}-\frac{\kappa^2}{2}\rho_m\delta=0.
\label{deleq_lin}
\ee

\begin{figure}[tb!]
\begin{center}
\includegraphics[angle=0, width=0.44
\textwidth]{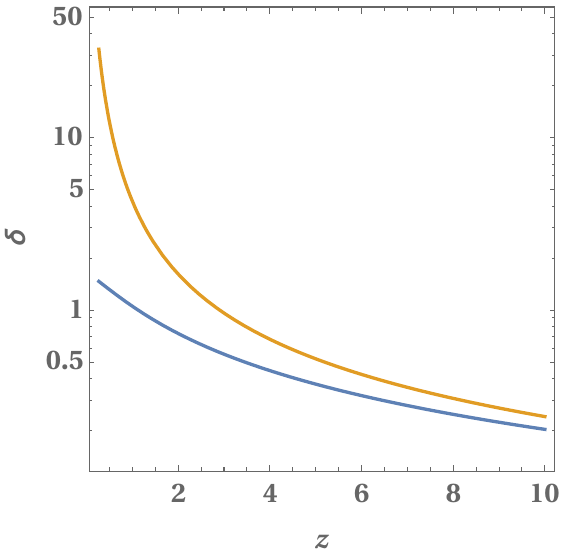}
\end{center}
\caption{{\small The evolution of matter density contrast in linear (blue curles) and non-linear (yellow curves) regime (for $p <1$ case) .}}
\label{densitycont}
\end{figure}

\noindent
Note that, in the above set of equations, we have ignored that fluctuations in the axion field. As mentioned before, the mass of the axion field is extremely small ($\sim 10^{-35}$ eV). Hence the axion field fluctuations can only play role in the horizon scales and beyond and one can safely ignore axion field fluctuations in the sub horizon scales.

\noindent
The evolution of matter density contrast at linear and non-linear regime are  shown in figure \ref{densitycont}. As the perturbation grows, the non-linear contributions in the evolution equation become significant. The quantity of interest of this collapse mechanism is the critical density $\delta_c$ defined as the value of linear density contrast at the time of collapse. The redshift ($z_c$) at which the spherical collapse of non-liner density contrast occurs, is changed with the change of the initial density contrast.

\begin{figure}[tb]
\begin{center}
\includegraphics[angle=0, width=0.4\textwidth]{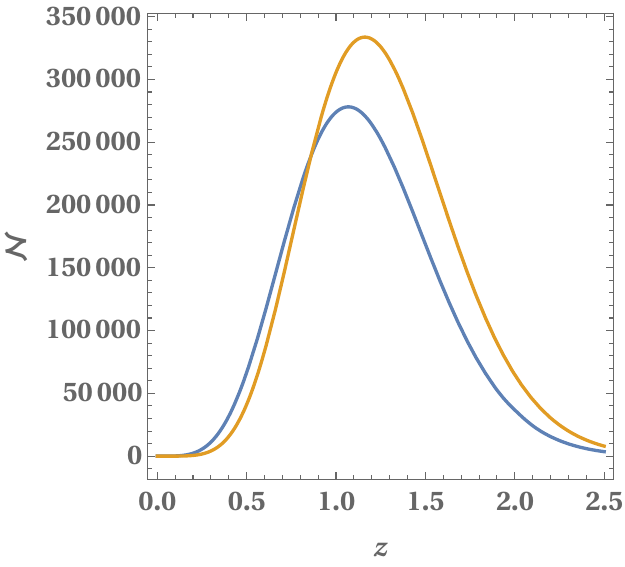}
\includegraphics[angle=0, width=0.4\textwidth]{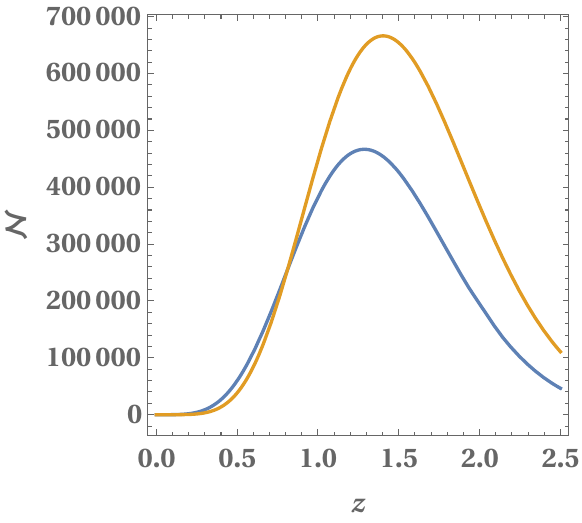}
\end{center}
\caption{{\small Distribution of galaxy cluster number count with redshift. The blue curves are obtained for the present model with parameter values kept at the best fit and the yellow curves are obtained for $\Lambda$CDM. The plots in the left panel are for the Press-Schechter fromalism of the mass function formula and the plots in the right panel are obtained for the Sheth-Tormen formalism of the mass function formula (for $p <1$ case) .}}
\label{Nplot_with_LCDM}
\end{figure}

\begin{figure}[tb]
\begin{center}
\includegraphics[angle=0, width=0.36\textwidth]{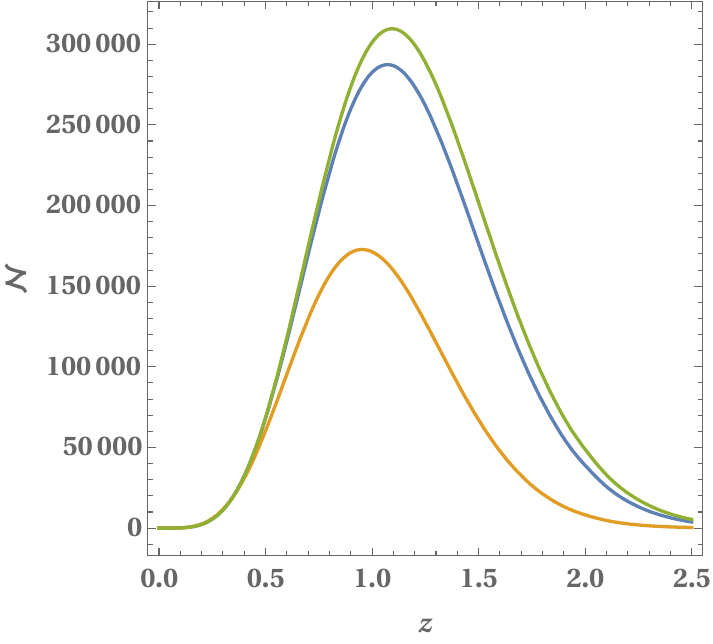}
\includegraphics[angle=0,width=0.36\textwidth]{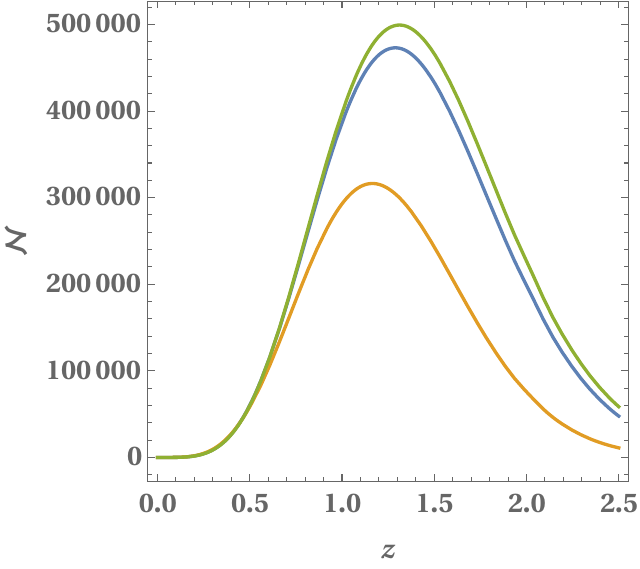}
\end{center}
\caption{{\small Distribution of cluster number count with redshift for different values the parameter $f_1$. The yellow curves is for $f_1=0.5 \sqrt{3}M_p$, blue curves are for $f_1=0.7219 \sqrt{3}M_p$ (the best fit value) and the green curve is for $f_1=0.9 \sqrt{3}M_p$. Other parameters are kept at the best fit values. The left panel is obtained for the Press-Schechter formalism and the right panel is obtained for the Sheth-Tormen formalism of the mass function (for $p <1$ case) .}}
\label{Nplot_varf1}
\end{figure}
\begin{figure}[tb]
\begin{center}
\includegraphics[angle=0, width=0.36\textwidth]{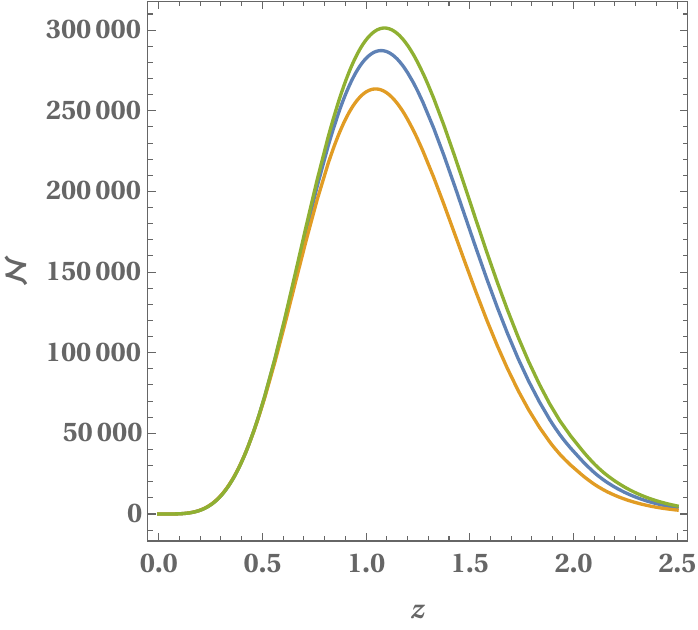}
\includegraphics[angle=0,width=0.36\textwidth]{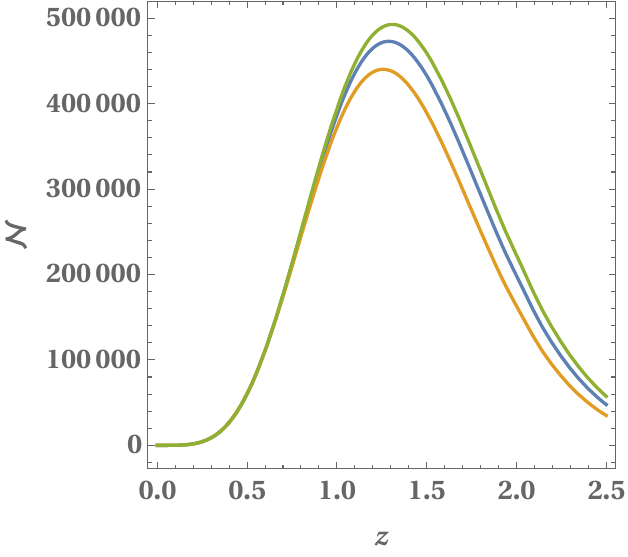}
\end{center}
\caption{{\small Distribution of cluster number count with redshift for different values of the parameter $p$. The yellow curves is for $p=0.3$, blue curves are for $p=0.5774$ (the best fit value) and the green curve is for $p=0.75$. Other parameters are fixed at the best fit values. The left panel is obtained for the Press-Schechter formalism and the right panel is obtained for the Sheth-Tormen formalism of the mass function formula (for $p <1$ case) .}}
\label{Nplot_vp}
\end{figure}

Next we shall study the distribution of the number density of collapsed object of a given mass range. There are two different mathematical formulation to evaluate the distribution of the number of collapsed objects along the redshift. The first one is the Press-Schechter formalism \cite{Press:1973iz} and the other one, which is a generalization of the first one, is called the Sheth-Tormen formalism \cite{Sheth:1999mn}. Both these formalism stand up with the assumption of a Gaussian distribution of the matter density field.  The comoving number density of collapsed object (galaxy clusters) at a certain redshift $z$ having mass range $M$ to $M+dM$ can be expressed as,

\be
\frac{dn(M,z)}{dM}=-\frac{\rho_{m0}}{M}\frac{d\ln{\sigma(M,z)}}{dM}f(\sigma(M,z)),
\ee
where $f(\sigma)$ is called the mass function. The mathematical formulation of the mass function was first proposed by Press and Schechter \cite{Press:1973iz}, which is given as,
\be
f_{PS}(\sigma)=\sqrt{\frac{2}{\pi}}\frac{\delta_c(z)}{\sigma(M,z)}\exp{\left[-\frac{\delta^2_c(z)}{2\sigma^2(M,z)} \right]}.
\ee
The $\sigma(M,z)$ is the corresponding rms density fluctuation in a sphere of radius $r$ enclosing a mass M. This can be expressed in terms of the linearised growth factor $g(z)=\delta(z)/\delta(0)$, and the rms of density fluctuation of at a distance $r_8=8h^{-1}$Mpc as,
\be
\sigma(z,M)=\sigma(0,M_8)\left(\frac{M}{M_8}\right)^{-\gamma/3}g(z),\ee
where $M_8=6\times10^{14}\Omega_{m0}h^{-1}M_{\odot}$, the mass within a sphere of radius $r_8$ and the $M_{\odot}$ is  the solar mass. The $\gamma$ is defined as
\be
\gamma=(0.3\Omega_{m0}h+0.2)\left[2.92+\frac{1}{3}\log{(\frac{M}{M_8})}\right].
\ee
Finally the effective number of collapsed objects between a mass range $M_i<M<M_s$ per redshift per square degree yield as,
\be
{\mathcal N}(z)=\int_{1deg^2}d\Omega\left( \frac{c}{H(z)}\left[\int_0^z\frac{c}{H(x)}dx\right]^2\right)\int_{M_i}^{M_s}\frac{dn}{dM}dM.
\ee
Though the Press-Shechter fromalism is successful to depict a general nature of the distribution of galaxy cluster number count, it suffers from the prediction of higher abundance of galaxy cluster at low redshift and lower abundance of clusters at high redshift compared to the result obtained in simulation of dark matter halo formation. To alleviate this issue, a modified mass function formula is proposed by Sheth and Tormen \cite{Sheth:1999mn}, which is given as,

\be
f_{ST}(\sigma)=A\sqrt{\frac{2}{\pi}}\left[1+\left( \frac{\sigma^2(M,z)}{a\delta^2_c(z)}\right)^p\right]\frac{\delta_c(z)}{\sigma(M,z)}\exp{\left[-\frac{a\delta_c^2(z)}{2\sigma^2(M,z)}\right]}.
\label{ST}
\ee
The Sheth-Tormen mass function formula, given in equation (\ref{ST}), introduce three new parameters $(a,p,A)$ and for the values $(1,0,\frac{1}{2})$ the Sheth-Torman mass function actually become the Press-Shechter mass function. In the present work, while studying the distribution of cluster number count using Sheth-Tormen mass function formula, the values of the parameter $(a,p,A)$ are fixed at $(0.707,0.3,0.322)$ as obtained form the simulation  of dark matter halo formation \cite{Reed:2006rw}. \\

 With this background, we study the influence of the axion field on the clustering of dark matter. We also study the effect of oscillations on the cluster number counts. To obtain the distribution of galaxy cluster number counts, the mass range considered is $10^{14}h^{-1}M_{\odot}<M<10^{16}h^{-1}M_{\odot}$ \cite{Nunes:2005fn}. In figure \ref{Nplot_with_LCDM} the cluster number count distribution with redshift are shown for the present model  of dark energy and also for the $\Lambda$CDM cosmology. We have shown the number count for both Press-Schechter and Sheth-Tormen formulation of the mass function. In this figure, the parameter values for axion are fixed at the best fit, given in section \ref{Obconstraint}, and for the $\Lambda$CDM model, the required parameter are fixed from the Planck 2018 results \cite{Aghanim:2018eyx}. It is observed that the axion model allows less number of clustered objects compared to the $\Lambda$CDM model. Next in Figure \ref{Nplot_varf1} and \ref{Nplot_vp}, we show the effect of variation of the parameter values on the number count distribution. In figure \ref{Nplot_varf1} the parameter $f_1$ is varied and the other parameters are kept fixed at the best fit values. The axion field decay constant ($f_1$) determines the frequency of the oscillation in the scalar field potential. A lower value of $f_1$ produces higher frequency of oscillation and vice versa. It is clear from figure \ref{Nplot_varf1} that the clustering of dark matter is highly suppressed with lower values of $f_{1}$ that increase in the frequency of the scalar field oscillation. It is also interesting to note that sub-Plancking values for $f_{1}$ suppresses the clustering in dark matter compared to the super-Planckian values for $f_{1}$. This is an interesting observation and has the potential to put strong constraints on $f_{1}$ from future large scale galaxy surveys.
 
 In figure \ref{Nplot_vp}, the parameter $p$ is varied and others are fixed at best the fit. The cluster number count is found to be less effected by the change in the parameter $p$.

\section{Conclusion}

The present study is related to a dark energy model involving axion field. The advantage of axion as dark energy is due to the fact that the flatness of the axion potential is always protected from quantum correction. On the other hand, to be consistent with the late time acceleration, the axion decay constant needs to be super-Planckian which is in conflict with the low-energy effective field theory of string compactification. To solve this problem, a multi-axion model in the presence of a cosmolgical constant (CC) has been proposed in String-Axiverse scenario \cite{Cicoli:2018kdo}.

In this present work, we study such scenario involving axion in the context of presently available cosmological data. We assume that the heavier fields in this multi-axion scenario, settle down to their minimum and the lightest field is still evolving to behave as a quintessence field. The effective CC in this scenario gets contribution from the original CC present in the model as well as from the ground state energy of the different axion field. We assume that asymptotic future is always de-Sitter, neglecting the possibility of this effective CC becomes negative at any time during the evolution.

With this, we confront the model with various observational data presently available, related to both background expansion as well as growth of structures. These are the important results in our study:

\begin{itemize}
   \item We show that cosmologically the two cases $\lambda_l^4 > \lambda^4$ and $\lambda_l^4 < \lambda^4$ are equivalent as they give similar constraints on the cosmological parameters.
 
    \item We show that sub-Planckian values for the axion decay constant $f_{1}$ in potential of equation 8 are allowed by the data. This ensures the validity of the low-energy effective field theory prescription of string compactification. This is one of the most important results in this study. 
    
    \item In terms of the different information criteria (AIC and BIC)  as well as  Bayesian evidences ( Table III and IV), the axion model is strongly favoured over $\Lambda$CDM model for both $p < 1$ and $p > 1$ cases.
    
    \item The results show that the scenario where the Universe is asymptotically anti de-Sitter (ADS) for $p>1$ is equally favoured compared to the scenario where the Universe is asymptotically de-Sitter (DS) for $p<1$. 
    
    \item At $2\sigma$ level, oscillatory behaviours in the scalar field evolution as well as in the dark energy equation state are consistent with the data in contrast to the equation of state in \cite{Dutta:2006cf} and \cite{Adak:2014moa}. This is allowed only because of the existence of the additional cosmological constant to the axion potential.
     
    \item
    The sound horizon at drag epoch $r_{d}$ is also within $2\sigma$ of Planck measured value.
    
    \item
    We show that the axion model  suppresses the cluster number counts compared to the $\Lambda$CDM model. This suppression is enhanced for sub-Planckian values of the axion decay constant $f_{1}$. Interestingly, the lower values of $f_{1}$ also result higher oscillations in the axion field evolution as well as in dark energy equation state. This is a very crucial observation (probably for the first time) related to the axions in String-Axiverse model, and can be smoking gun to probe such models with future large scale galaxy surveys. Moreover, our results show that a Universe with an overall negative cosmological constant is well consistent with cosmological data which may have far reaching consequences for dark energy model building in String theory.
    
  \end{itemize}
    
    To conclude, we do a detail study related to observational constraints on axion as dark energy in the String-Axiverse scenario. The results from this work can have far reaching consequences for axion models as a candidate for dark energy in near future.

\vskip 1.5 cm

\noindent
{\bf Acknowledgment:} Ruchika acknowledges the CSIR, Govt of India for funding under Senior Research Fellowship. AAS acknowledges funding from DST-SERB, Govt of India, under the project NO. MTR/20l9/000599. The work of KD was partially supported by the grant MTR/2019/000395, funded by SERB, DST, Government of India. AM acknowledges the financial support from the Science and Engineering Research Board (SERB), Department of Science and Technology, Government of India as a National Post-Doctoral Fellow (NPDF, File no. PDF/2018/001859). Ruchika and AAS would like to thank Theory Division of Saha Institute of Nuclear Physics, Kolkata for academic visits when the project was initiated. The authors also acknowledge the use of High Performance Computing facility Pegasus at IUCAA, Pune, India.

\end{document}